\documentclass[12pt]{article}
\usepackage{enumerate}
\usepackage{amsfonts}
\usepackage{amsmath}
\usepackage{amssymb}
\usepackage{amsthm}

\def\D{\mathcal{D}}

\def\A{\mathcal{A}}

\def\H{\mathcal{H}}
\def\K{\mathcal{K}}

\def\S{\mathfrak{S}}

\def\T{\mathfrak{T}}
\def\B{\mathfrak{B}}

\newcommand{\rank}{\mathrm{rank}}

\newcommand{\id}{\mathrm{Id}}
\newcommand{\Tr}{\mathrm{Tr}}

\newcommand{\shs}{\hspace{1pt}}
\newcommand{\sn}{\|\hspace{-1pt}|}

\newcounter{defin}  \newcounter{lemma}  \newcounter{theorem}
\newcounter{property} \newcounter{corol}  \newcounter{remark} \newcounter{example}

\newenvironment{lemma}{\par\refstepcounter{lemma}%\noindent
     \textbf{Lemma \thelemma.} }{\rm\par}
\newenvironment{theorem}{\par\refstepcounter{theorem}%\noindent
     \textbf{Theorem \thetheorem.}\ }{\rm\par}
\newenvironment{property}{\par\refstepcounter{property}%\noindent
     \textbf{Proposition \theproperty.}\ }{\rm\par}

\newenvironment{definition}{\par\refstepcounter{defin}%\noindent
     \textbf{Definition \thedefin.}\ }{\rm\par}
\newenvironment{remark}{\par\refstepcounter{remark}%\noindent
     \textbf{Remark \theremark.}}{\rm\par}

\begin{document}

\title{On extension of quantum channels and operations to the space of relatively bounded operators}

\author{M.E. Shirokov\footnote{Steklov Mathematical Institute, RAS, Moscow, email:msh@mi.ras.ru}}
\date{}
\maketitle

\vspace{-10pt}

\begin{abstract}
We analyse possibility to extend a quantum operation (sub-unital normal CP linear map on the algebra  $\B(\H)$  of bounded operators on a separable Hilbert space $\H$) to the space of all
operators on $\H$ relatively bounded w.r.t. a given positive unbounded operator.

We show that a quantum operation $\,\Phi\,$ can be uniquely extended to a bounded linear operator
on the Banach space of all $\sqrt{G}$-bounded operators on $\H$ provided that
the operation $\Phi$ is $G$-limited:  the predual operation $\Phi_*$ maps the set of positive trace class operators $\rho$ with finite $\mathrm{Tr}\rho G$ into itself.

Assuming that $G$ has discrete spectrum of finite multiplicity we prove that for a wide class of quantum operations the existence of the above extension implies the $G$-limited property.

Applications to the theory of Bosonic Gaussian channels are considered.

%The converse statement is proved for all channels and operations with finite Choi rank provided that the operator $G$ has discrete spectrum of finite multiplicity.
\end{abstract}

\section{Introduction}

Unital and subunital completely positive (CP) normal linear maps between algebras of bounded operators play important role in the quantum theory.
Unital CP linear maps called \emph{quantum channels} describe evolution of open quantum systems (in the Heisenberg picture),
subunital CP linear maps called {\emph{quantum operations} are also used essentially, in particular, in the theory of quantum measurements \cite{H-SCI,Watrous,Wilde}.

In the standard description of a quantum system dynamics a quantum channel (operation) acts on bounded observables -- Hermitian bounded operators on a separable Hilbert space associated with the system. But for many important physical quantities (the position, momentum, etc.) the corresponding observables are
unbounded Hermitian operators. So, the question arises of extending the quantum channel (operation) to unbounded observables. For a given channel (operation)\footnote{$\B(\H)$ is the  algebra of all bounded operators on a Hilbert space $\H$.} $\Phi:\B(\H)\rightarrow\B(\H)$ this question can be solved by using the Stinespring representation
\begin{equation}\label{St-rep-}
\Phi(A)=V^*_{\Phi}(A\otimes I_{\K})V_{\Phi}
\end{equation}
where $V_{\Phi}$ is an isometry (contraction) from the system  space $\H$ into the tensor product of this space and some separable Hilbert space $\K$ (typically called the environment space). Indeed, for any Hermitian unbounded operator $A$ the r.h.s. of (\ref{St-rep-}) defines some linear operator on $\H$. But we can say nothing  about the domain of this operator, since in general the range of $V_{\Phi}$ does not coincide with $\H\otimes\K$.
In particular, we can not assert that the domain of this operator is dense in $\H$.\footnote{It is easy to construct
a channel $\Phi$ and a positive operator $A$ such that the domain of the r.h.s. of (\ref{St-rep-}) contains only the zero vector.}

For quantum channels (operations) of special classes the existence of an adequate extension to unbounded operators are well known. For example, any Bosonic Gaussian channel is well defined on the algebra of all polynomials of canonical observables (which are unbounded operators) and transforms this algebra into itself \cite[Ch.12]{H-SCI}.
In this paper we analyze the extension problem in full generality not assuming in advance any special properties of a quantum channel (operation).

Speaking about extension of quantum channels (operations) to unbounded operators we will restrict attention to operators relatively bounded
w.r.t. the operator $\sqrt{G}$ (briefly $\sqrt{G}$-bounded operators), where $G$ is a given positive densely defined operator on $\H$, i.e. to the operators $A$ well defined on the domain
$\mathcal{D}(\sqrt{G})$ of $\sqrt{G}$ such that
\begin{equation*}%\label{rb-rel}
\|A\varphi\|^2\leq a^2\|\varphi\|^2+b^2\|\sqrt{G}\varphi\|^2,\quad \forall \varphi\in\mathcal{D}(\sqrt{G}),
\end{equation*}
for some  nonnegative numbers $a$ and $b$ \cite{Kato,BS}. This is explained, in particular, by the following physical reason: the quantum observables corresponding to
important physical quantities are $\sqrt{G}$-bounded
operators provided that either $G$ or some power of $G$ is a Hamiltonian of a quantum system.

The linear space of all $\sqrt{G}$-bounded operators can be made a Banach space by equipping it with any of the operator
\emph{E}-norms $\|\cdot\|_E^G$ induced by $G$ \cite{ECN}.\footnote{The advantages of the norm $\|\cdot\|_E^G$ in comparison with the equivalent norm commonly used on  set of relatively bounded operators are described in Section 2.2.} For given $E>0$  the operator
\emph{E}-norm of a bounded operator $A$ is defined as
\begin{equation*}%\label{ec-on}
 \|A\|^{G}_E\doteq \sup_{\substack{\rho\in\mathfrak{S}(\mathcal{H}):
\Tr G\rho \leq E}}\sqrt{\Tr A\rho A^*},
\end{equation*}
where the supremum is over all states $\rho$ (positive trace class operators on $\H$ with unit trace) such that  $\Tr G\rho \leq E$.
This definition is naturally extended to all $\sqrt{G}$-bounded operators (see Section 2.2).
\smallskip

In this paper we prove that  formula (\ref{St-rep-}) defines a bounded linear operator on the Banach space $\B_G(\H)$ of all
$\sqrt{G}$-bounded operators with the norm $\|\cdot\|_E^G$ provided that
\begin{equation}\label{E-phi-}
\Tr G\Phi_*(\rho)<+\infty\;\textrm{ for any state}\;\rho\;\textrm{ such that }\;\Tr G\rho<+\infty,
\end{equation}
where $\Phi_*$ is the predual map to the channel (operation) $\Phi$. We also prove that for a large class of quantum channels and operations
condition (\ref{E-phi-}) is necessary for continuity of the map $A\mapsto\Phi(A)$ on $\B(\H)$ w.r.t. the norm $\|\cdot\|_E^G$.
So, if $\Phi$ is a quantum channel (operation)  from this class then formula (\ref{St-rep-}) defines a bounded linear operator on the Banach space $\B_G(\H)$ \emph{if and only if} condition (\ref{E-phi-}) holds.

By noting that any Bosonic Gaussian channel $\Phi$ satisfies condition (\ref{E-phi-}) provided that $G$ is the number operator in the multimode Bosonic system we consider some applications of the above general results to these channels.

\section{Preliminaries}

\subsection{Basic notations}

Let $\mathcal{H}$ be a separable infinite-dimensional Hilbert space, $\mathfrak{B}(\mathcal{H})$
-- the algebra of all bounded operators on $\mathcal{H}$ with the operator norm $\|\!\cdot\!\|$ and $\mathfrak{T}(\mathcal{H})$ --
the Banach space of all trace-class operators on $\mathcal{H}$ with the trace norm $\|\!\cdot\!\|_1$ (the Schatten class of order 1) \cite{Kato,BS}. Let
$\mathfrak{T}_{+}(\mathcal{H})$ be the positive cone in $\mathfrak{T}(\mathcal{H})$ and $\mathfrak{S}(\mathcal{H})$ the set of quantum states --  operators in
$\mathfrak{T}_{+}(\mathcal{H})$ with unit trace \cite{H-SCI}.

Denote by $I_{\H}$ the unit operator on a Hilbert space
$\mathcal{H}$ and by $\id_{\mathcal{\H}}$ the identity
transformation of the Banach space $\mathfrak{T}(\mathcal{H})$.

A completely positive (CP) normal unital (corresp., subunital) linear map $\,\Phi:\B(\H)\rightarrow \B(\H)\,$ is called \emph{quantum channel (corresp., operation) in the Heisenberg picture}.\footnote{The map $\Phi$ is called normal if $\Phi(\sup_{\lambda}A_\lambda)=\sup_{\lambda}\Phi(A_{\lambda})$ for any increasing net $A_\lambda\subset\B(\H)$ \cite{B&R,H-SSQT}. This property is equivalent to existence of the predual map  $\,\Phi_*:\T(\H)\rightarrow \T(\H)\,$. The map $\Phi$ is called unital (corresp., subunital) if $\Phi(I_{\H})=I_{\H}$ (corresp., $\Phi(I_{\H})\leq I_{\H}$).} Its predual map  $\,\Phi_*:\T(\H)\rightarrow \T(\H)\,$ defined by the relation
$$
\Tr \Phi_*(\rho)A=\Tr \Phi(A)\rho,\quad A\in\B(\H),\;\rho\in\T(\H),
$$
is called \emph{quantum channel (corresp., operation) in the Schrodinger picture} \cite{H-SSQT}.

For any quantum channel (corresp., operation) $\Phi$ the Stinespring theorem (cf.\cite{St}) implies existence of a separable Hilbert space
$\mathcal{K}$ and an isometry (corresp., contraction)
$V_{\Phi}:\mathcal{H}\rightarrow\mathcal{H}\otimes\mathcal{K}$ such
that
\begin{equation}\label{St-rep}
\Phi(A)=V^*_{\Phi}(A\otimes I_{\K})V_{\Phi},\quad
A\in\mathfrak{B}(\mathcal{H}),
\end{equation}
and, respectively,
\begin{equation*}%\label{St-rep}
\Phi_*(\rho)=\mathrm{Tr}_{\K}V_{\Phi}\rho V_{\Phi}^{*},\quad
\rho\in\mathfrak{T}(\mathcal{H}),
\end{equation*}
where $\Tr_{\K}$ denotes the partial trace over the space $\K$. The Stinespring representation (\ref{St-rep}) is called \emph{minimal} if
the linear span of the set $\,\{(A\otimes I_{\K}) V_{\Phi}\varphi\,|\, A\in\B(\H), \varphi\in\H \}\,$\\  is dense in $\H\otimes\K$. The dimension of the space $\K$ in the minimal Stinespring representation is called the \emph{Choi rank} of a channel (operation) \cite[Ch.6]{H-SCI}. \smallskip

For any quantum operation $\,\Phi:\B(\H)\rightarrow \B(\H)\,$ the following inequality (called Kadison's inequality)
\begin{equation}\label{Kad}
 \Phi(A^*)\Phi(A)\leq \|\Phi(I_{\H})\|\Phi(A^*A)
\end{equation}
holds for any $A\in\B(\H)$.

We will consider  unbounded densely defined positive operators on $\H$ having discrete spectrum of finite multiplicity.
In Dirac's notations any such operator ${G}$ can be represented as
\begin{equation}\label{H-rep}
G=\sum_{k=0}^{+\infty} E_k |\tau_k\rangle\langle\tau_k|
\end{equation}
and the domain $\mathcal{D}(G)=\{ \varphi\in\H\,|\,\sum_{k=0}^{+\infty} E^2_k|\langle\tau_k|\varphi\rangle|^2<+\infty\}$, where
$\left\{\tau_k\right\}_{k=0}^{+\infty}$ is the orthonormal
basis of eigenvectors of ${G}$ corresponding to the nondecreasing sequence $\left\{\smash{E_k}\right\}_{k=0}^{+\infty}$ of eigenvalues
tending to $+\infty$. We will use the following (cf.\cite{W-EBN})\smallskip

\begin{definition}\label{D-H}
An operator ${G}$ having representation (\ref{H-rep}) is called \emph{discrete}.
\end{definition}\smallskip

\subsection{Relatively bounded operators and the operator \emph{E}-norms}

In this paper we will consider  operators on a separable Hilbert space $\H$ relatively bounded with
respect to a given positive semidefinite densely defined operator. For our purposes it is convenient to denote this positive
operator by $\sqrt{G}$ assuming that $G$ is a positive semidefinite operator on $\H$ with  dense domain $\mathcal{D}({G})$ such that
\begin{equation}\label{G-cond}
\inf\left\{\shs\|G\varphi\|\,|\,\varphi\in\mathcal{D}({G}),\|\varphi\|=1\shs\right\}=0.
\end{equation}

 A linear operator $A$ is called relatively bounded w.r.t. the operator $\sqrt{G}$ (briefly, $\sqrt{G}$-bounded) if
$\mathcal{D}(\sqrt{G})\subseteq \D(A)$ and
\begin{equation}\label{rb-rel}
\|A\varphi\|^2\leq a^2\|\varphi\|^2+b^2\|\sqrt{G}\varphi\|^2,\quad \forall \varphi\in\mathcal{D}(\sqrt{G}),
\end{equation}
for some  nonnegative numbers $a$ and $b$ \cite{Kato}. The $\sqrt{G}$-bound of $A$ (denoted by $b_{\sqrt{G}}(A)$ in what follows) is defined as
the infimum of the values $b$ for which (\ref{rb-rel}) holds with some $a$. If the $\sqrt{G}$-bound is equal to zero then $A$ is called $\sqrt{G}$-infinitesimal operator (infinitesimally bounded w.r.t. $\sqrt{G}$) \cite{Kato,R&S+,BS}.

Since $\sqrt{G}$ is a closed operator, the linear space $\D(\sqrt{G})$ equipped with the inner product
$$
\langle\varphi|\psi\rangle^G_E=\langle\varphi|\psi\rangle+\langle\varphi|G|\psi\rangle/E,\quad E>0,
$$
is a Hilbert space denoted by $\H_E^G$ in what follows \cite{R&S}. A restriction of any $\sqrt{G}$-bounded operator to the set $\D(\sqrt{G})$ can be treated as
bounded operator from $\H_E^G$ into $\H$ and, vise versa, any bounded operator from $\H_E^G$ into $\H$ induces a $\sqrt{G}$-bounded operator on $\H$.
Thus, the linear space of all $\sqrt{G}$-bounded operators on $\H$ equipped with the norm
\begin{equation}\label{eq-norms-2}
\sn A\sn^{G}_{E}=\sup_{ \varphi\in \D(\sqrt{G})}\frac{\|A\varphi\|}{\sqrt{\|\varphi\|^2+\|\sqrt{G}\varphi\shs\|^2/E}}
\end{equation}
is a Banach space.\footnote{We identify operators coinciding on $\D(\sqrt{G})$.} For our purposes it is more convenient to  use the equivalent norm
\begin{equation}\label{ec-on}
 \|A\|^{G}_E\doteq \sup_{\substack{\rho\in\mathfrak{S}(\mathcal{H}):
\Tr G\rho \leq E}}\sqrt{\Tr A\rho A^*}
\end{equation}
on the linear space of all $\sqrt{G}$-bounded operators. The supremum here is over all states $\rho$ in $\S(\H)$ such that  $\Tr G\rho\leq E$.\footnote{The value of $\Tr {G}\rho$ (finite or infinite) is defined as $\sup_n\Tr P_nG\rho$, where $P_n$ is the spectral projector of $G$ corresponding to the interval $[0,n]$.} By Lemma 5 in \cite{ECN} for any $\sqrt{G}$-bounded operator $A$ the function $\rho\mapsto A\rho A^*$ is well defined on the set $\,\T^+_{G}\doteq \{\rho\in\T_{+}(\H)\,|\,\Tr {G}\rho<+\infty\shs\}\,$ by the expression
\begin{equation}\label{aa-d}
  A\rho A^*\doteq\sum_i|\alpha_i\rangle\langle\alpha_i|,\qquad |\alpha_i\rangle=A|\varphi_i\rangle,
\end{equation}
where $\rho=\sum_i |\varphi_i\rangle\langle\varphi_i|$ is any decomposition of $\rho\in\T^+_{G}$ into $1$-rank positive operators. This function is affine
and takes values in $\T_{+}(\H)$. So, the r.h.s. of (\ref{ec-on}) is well defined for any $\sqrt{G}$-bounded operator $A$. Due to condition (\ref{G-cond})
the supremum in (\ref{ec-on}) can be taken over all operators $\rho\in\T_+(\H)$ such that $\Tr G\rho\leq E$ and $\Tr \rho\leq 1$ \cite[Prop.3]{ECN}.\smallskip

The norm $\|\cdot\|^{G}_{E}$ called the operator \emph{E}-norm  in \cite{ECN} can be also defined by the following equivalent expressions
\begin{equation}\label{en-def-r-1}
 \!\!\textstyle \|A\|^{G}_E=\sup\left\{\sqrt{\sum_i\|A\varphi_i\|^2}\,\left|\,\{\varphi_i\}\subset\D(\sqrt{G}):\sum_i\|\varphi_i\|^2\leq 1,\;\sum_i\|\sqrt{G}\varphi_i\|^2\leq E\right.\right\}
\end{equation}
and
\begin{equation}\label{en-def-r-2}
\textstyle \|A\|^{G}_E=\sup\left\{\|A\otimes I_{\K}\varphi\|\,\left|\,\varphi\in\D(\sqrt{G}\otimes I_{\K}):\|\varphi\|\leq 1,\;\|\sqrt{G}\otimes I_{\K}\varphi\|^2\leq E\right.\right\},
\end{equation}
where $\K$ is a separable infinite-dimensional Hilbert space. If $G$ is a discrete  operator (Def.\ref{D-H}) then all the above expressions are simplified as follows
\begin{equation}\label{en-def-r-1++}
 \textstyle \|A\|^{G}_E=\sup\left\{\|A\varphi\|\,\left|\,\varphi\in\D(\sqrt{G}):\|\varphi\|\leq 1,\;\|\sqrt{G}\varphi\|^2\leq E\right.\right\}.
\end{equation}
Validity of the simplified expression (\ref{en-def-r-1++}) in the case of arbitrary positive operator $G$ is an interesting open question (see the Appendix in \cite{ECN}).\footnote{The r.h.s. of (\ref{en-def-r-1++}) defines  a norm on the set of all $\sqrt{G}$-bounded operators denoted by $\|\cdot\|^{G}_{\circ,E}$ in \cite{ECN}, where it is shown that this norm is equivalent to the norms $\|\cdot\|^{G}_{E}$ and $\sn \cdot\sn^{G}_{E}$ and that the function $E\mapsto\left[\|A\|^{G}_{E}\right]^2$ is the concave envelope (hull) of the function $E\mapsto\left[\|A\|^{G}_{\circ,E}\right]^2$. It means, in particular, that the conjectured coincidence of $\|\cdot\|^{G}_{E}$ and $\|\cdot\|^{G}_{\circ,E}$ is equivalent to concavity of the function   $E\mapsto\left[\|A\|^{G}_{\circ,E}\right]^2$ for any $\sqrt{G}$-bounded operator $A$.}

For any $\sqrt{G}$-bounded operator $A$ both norms $\|A\|^{G}_E$ and $\sn A\sn^{G}_E$ are nondecreasing functions of $E$ tending to $\|A\|\leq+\infty$ as $E\rightarrow+\infty$. They are related by the inequalities
\begin{equation}\label{eq-one}
\sqrt{1/2}\|A\|^{G}_{E}\leq \sn A\sn^{G}_{E}\leq\|A\|^{G}_{E},
\end{equation}
which show the equivalence of these norms on the
set of all $\sqrt{G}$-bounded operators \cite{ECN}. Moreover, for any $\sqrt{G}$-bounded operator $A$ the functions
$E\mapsto \|A\|^{G}_{E}$ and $E\mapsto \sn A\sn^{G}_{E}$ are completely determined by each other via the following expressions (\cite[Th.3A]{ECN}):
$$
\sn A\sn^{G}_{E}=\sup_{t>0}\|A\|^{G}_{tE}/\sqrt{1+t},\quad  \|A\|^{G}_{E}=\inf_{t>0}\sn A \sn^{G}_{tE}\sqrt{1+1/t},\quad E>0.
$$

One of the main advantages of the norm $\|A\|^{G}_E$ is the concavity of the function $E\mapsto\left[\|A\|^{G}_{E}\right]^p$ for any $p\in(0,2]$ and any $\sqrt{G}$-bounded operator $A$ which essentially simplifies quantitative analysis of functions depending on $\sqrt{G}$-bounded operators \cite[Section 5]{ECN}.\footnote{The function $E\mapsto\left[\sn A\sn^{G}_{E}\right]^p$ is not concave in general for any $p\in(0,2]$ \cite[Section 3.1]{ECN}.}
This property implies, in particular, that
\begin{equation}\label{E-n-eq}
\|A\|^{G}_{E_1}\leq \|A\|^{G}_{E_2}\leq \sqrt{E_2/E_1}\|A\|^{G}_{E_1}\quad\textrm{ for any } E_2>E_1>0.
\end{equation}
Hence for given operator ${G}$ all the norms $\|\!\cdot\!\|^{G}_{E}$, $E>0$, are equivalent on the
set of all $\sqrt{G}$-bounded operators. By inequalities (\ref{eq-one})
the same is true for the norms $\sn\!\cdot\!\sn^{G}_{E}$, $E>0$.

Another advantage of the norm $\|A\|^{G}_E$  essentially used in this paper is the possibility to estimates the norms $\|\Phi(A)\|^{G}_E$
via the norm $\|A\|^{G}_E$ of any bounded operator $A$, where $\Phi$ is a $2$-positive linear transformation of $\B(\H)$ satisfying the particular conditions \cite[Proposition 5E]{ECN}.

Denote by $\B_G(\H)$ the linear space of all $\sqrt{G}$-bounded operators equipped with any of the equivalent norms $\|A\|^{G}_E$, $E>0$. The equivalence of the norms $\|A\|^{G}_E$ and $\sn A\sn^{G}_E$ mentioned before implies that $\B_G(\H)$ is a (nonseparable) Banach space. The\break $\sqrt{G}$-bound $b_{\sqrt{G}}(\cdot)$ is a continuous seminorm on $\B_G(\H)$, for any operator $A\in\B_G(\H)$ it can be determined by the formula
\begin{equation}\label{G-bound}
b_{\sqrt{G}}(A)=\lim_{E\rightarrow+\infty}\|A\|^{G}_{E}/\sqrt{E},
\end{equation}
where the limit can be replaced by infimum over all $E>0$ \cite[Theorem 3B]{ECN}.\smallskip

The closed subspace $\,\B^0_{\!G}(\H)$ of $\,\B_{\!G}(\H)$ consisting of all $\sqrt{G}$-infinitesimal operators, i.e. operators with
the $\sqrt{G}$-bound equal to $0$, coincides with the completion of $\B(\H)$ w.r.t. any of the norms $\|\!\cdot\!\|^{G}_E$, $E>0$ \cite[Theorem 3C]{ECN}.\smallskip

We will use the following  observation \cite[Lemma 4, Theorem 3A]{ECN}.\smallskip

\begin{lemma}\label{vbl} \emph{If $A$ is a $\sqrt{G}$-bounded operator on $\H$ then for any separable Hilbert space $\K$  the operator $A\otimes I_{\K}$ naturally defined on the set $\,\D(\sqrt{G})\otimes\K$  has a unique linear $\sqrt{G}\otimes I_{\K}$-bounded extension to the set $\,\D(\sqrt{G}\otimes I_{\K})$.\footnote{$\D(\sqrt{G})\otimes \K$ is the linear span of all the vectors $\varphi\otimes\psi$, where $\varphi\in\D(\sqrt{G})$ and $\psi\in\K$.} This extension (also denoted by $A\otimes I_{\K}$) has the following property
\begin{equation}\label{s-prop}
 A\otimes I_{\K}\!\left(\sum_{i}|\varphi_i\rangle\otimes|\psi_i\rangle\right)=\sum_{i}A|\varphi_i\rangle\otimes|\psi_i\rangle
\end{equation}
for any countable sets $\{\varphi_i\}\subset\D(\sqrt{G})$ and $\{\psi_i\}\subset\K$ such that $\sum_{i}\|\sqrt{G}\varphi_i\|^2<+\infty$ and $\langle\psi_i|\psi_j\rangle=\delta_{ij}$, which implies that $\|A\otimes I_{\K}\|^{{G}\otimes I_{\K}}_E=\|A\|^{{G}}_E$ for any $E>0$.}
\end{lemma}\medskip

\begin{remark}\label{w-c} Property (\ref{s-prop}) implies that
$$
(A\otimes I_{\K})(I_{\H}\otimes W)|\varphi\rangle=(I_{\H}\otimes W)(A\otimes I_{\K})|\varphi\rangle
$$
for any $\varphi\in\D(\sqrt{G}\otimes I_{\K})$ and a  partial isometry $W\in\B(\K)$ s.t.
$I_{\H}\otimes W^{*}W |\varphi\rangle=|\varphi\rangle$.
\end{remark}

\subsection{On $G$-limited quantum operations}

Let $\Phi:\B(\H)\rightarrow\B(\H)$ be a unital CP normal linear map (quantum channel in the Heisenberg picture) and
$\Phi_*:\T(\H)\rightarrow\T(\H)$ its predual map (quantum channel in the Schrodinger picture). Let ${G}$ be a positive (semidefinite) operator on $\H$ with dense domain $\mathcal{D}({G})$ satisfying condition (\ref{G-cond}). If $G$ is treated as a Hamiltonian of a quantum system associated with the space $\H$ then for any state $\rho$ in $\S(\H)$ the value of $\Tr G\rho$ (finite or infinite) is the (mean) energy of (the system in) the state $\rho$ \cite{H-SCI}. The states with finite energy can be produced (in principal) in a physical experiment while
the states with infinite energy are physically unrealizable. So, it is reasonable to assume that a real (physically realizable) quantum channel $\Phi_*$ maps
states with finite energy into states with finite energy, i.e.
\begin{equation}\label{E-phi+}
\Tr G\Phi_*(\rho)<+\infty\;\textrm{ for any state}\;\rho\;\textrm{ such that }\;\Tr G\rho<+\infty.
\end{equation}

A quantum channel $\Phi$ satisfying the following formally stronger condition
\begin{equation}\label{E-phi}
Y_{\Phi}(E)\doteq \sup\left\{\shs\Tr {G}\Phi_*(\rho)\,|\,\rho\in\mathfrak{S}(\mathcal{H}),\Tr {G}\rho\leq E\,\right\}<+\infty
\end{equation}
for some $E>0$ is called energy-limited in \cite{W-EBN}, where it is pointed that this condition holds for many quantum channels used in applications. Due to condition (\ref{G-cond}) the function $E\mapsto Y_{\Phi}(E)$ is concave on $\mathbb{R}_+$. So, if $Y_{\Phi}(E)$ is finite for some $E>0$ then it is finite for all $E>0$. The quantity $Y_{\Phi}(E)/E$ can be called  the energy amplification factor of a quantum channel (operation) $\Phi$.\smallskip

In fact, conditions (\ref{E-phi+}) and (\ref{E-phi}) are equivalent. This follows from Lemma \ref{sl} below.\smallskip

To not be limited to the case when $G$ is a Hamiltonian of a quantum system we will use the following \smallskip

\begin{definition}\label{ELC} A quantum operation\footnote{i.e. sub-unital CP normal linear map} $\Phi:\B(\H)\rightarrow\B(\H)$ is called $G$-limited if
equivalent conditions (\ref{E-phi+}) and (\ref{E-phi}) hold.
\end{definition}\smallskip

The structure of $\S(\H)$ as a convex set implies the following useful fact.\smallskip

\begin{property}\label{el-p}
\emph{If  $G$ is a positive unbounded discrete operator (\ref{H-rep}) then
$$
Y_{\Phi}(E)=\sup\left\{\shs\Tr {G}\Phi_*(|\varphi\rangle\langle\varphi|)\,\left|\, \varphi\in\H_*, \|\sqrt{G}\varphi\|^2\leq E, \|\varphi\|=1\!\right.\right\}
$$
for any quantum operation $\,\Phi:\B(\H)\rightarrow\B(\H)$, where $\H_*$ is the linear span of all the eigenvectors $\,\tau_1,\tau_2,...\,$ of $G$, i.e. the supremum in (\ref{E-phi})
can be taken only over pure states corresponding to the vectors in $\H_*$.}
\end{property}\smallskip

Proposition \ref{el-p} follows from Lemma \ref{nbl} below, since the functions $\rho\mapsto\Tr {G}\rho$
and $\rho\mapsto\Tr {G}\Phi_*(\rho)$ are affine and lower semicontinuous.\smallskip

\begin{lemma}\label{nbl}
\emph{Let $G$ be a positive unbounded discrete operator (\ref{H-rep}) on a Hilbert space $\H$ and  $f$ a convex lower semicontinuous function on $\S(\H)$.
Then
$$
\sup\left\{\,f(\rho)\,|\, \rho\in\S(\H),\Tr G\rho\leq E\,\right\}=\sup\left\{f(|\varphi\rangle\langle\varphi|)\left|\, \varphi\in\H_*, \|\sqrt{G}\varphi\|^2\leq E, \|\varphi\|=1\!\right.\right\},
$$
where $\H_*$ is the linear span of all the eigenvectors $\,\tau_1,\tau_2,...\,$ of $\,G$.}
\end{lemma}\smallskip

\emph{Proof.} Denote by $\H_n$ and $P_n$ the subspace spanned by the eigenvectors $\,\tau_1,\tau_2,...,\tau_{n-1}\,$ and the projector
on this subspace correspondingly.

Assume that $A=\sup\left\{\,f(\rho)\,|\, \rho\in\S(\H),\Tr G\rho\leq E\,\right\}$ is finite.  For given arbitrary $\varepsilon>0$ let $\rho_{\varepsilon}$ be a state in $\S(\H)$ such that $\Tr G\rho_{\varepsilon}\leq E$ and $f(\rho_{\varepsilon})>A-\varepsilon$.
For each $n$ let $\rho_n=[\Tr P_n\rho_{\varepsilon}]^{-1}P_n\rho_{\varepsilon}P_n$. It is easy to see that
$\Tr G\rho_n\leq E$ provided that $E_n\geq E$ and that the sequence $\{\rho_n\}$ tends to $\rho_{\varepsilon}$.
By the lower semicontinuity of $f$ we have
$$
\liminf_{n\rightarrow+\infty} f(\rho_n)\geq f(\rho_{\varepsilon})>A-\varepsilon.
$$
It follows that there is $m$ such that $f(\rho_m)> A-2\varepsilon$. Since $\varepsilon$ is arbitrary, this shows that
\begin{equation}\label{A-exp}
A=\sup_n \sup\left\{\,f(\rho)\,|\, \rho\in\S(\H_n),\Tr G\rho\leq E\,\right\}.
\end{equation}
By using Lemma 2 in \cite{ECN} and the convexity  of $f$ we obtain
$$
\sup\left\{\,f(\rho)\,|\, \rho\in\S(\H_n),\Tr G\rho\leq E\,\right\}=\sup\left\{f(|\varphi\rangle\langle\varphi|)\left| \varphi\in\H_n, \|\sqrt{G}\varphi\|^2\leq E, \|\varphi\|=1\!\right.\right\}.
$$
This and (\ref{A-exp}) imply the assertion of the lemma, since $\H_*=\bigcup_{n}\H_n$.
\smallskip

The case $A=+\infty$ is considered similarly. $\square$ \smallskip

\begin{lemma}\label{sl}
\emph{Any  finite concave nonnegative function on a closed convex subset of a Banach space is bounded on this subset.}
\end{lemma}\smallskip

\emph{Proof.} Let $f$ be a finite concave nonnegative function on a closed convex subset $X$ of a Banach space.
Assume that for each $n\in \mathbb{N}$ there is $\,x_n\in X$ such that $f(x_n)>2^n$. Let $x_*=\sum_{n=1}^{+\infty}2^{-n}x_n$ and
$x^m_*=(1-p_m)^{-1}\sum_{n=1}^{m}2^{-n}x_n$, where $p_m=\sum_{n>m}2^{-n}$, $m\in \mathbb{N}$. By the concavity and nonnegativity of the function $f$ we have
$$
f(x_*)\geq (1-p_m)f(x_*^m)\geq \sum_{n=1}^{m} 2^{-n}f(x_n)> m
$$
for any $m$. This implies that $f(x_*)=+\infty$. $\square$

\section{The main results}

Part A of the following theorem gives a sufficient condition for existence of a continuous linear extension
of a quantum channel (operation) to the Banach space $\B_{\!G}(\H)$ of\break $\sqrt{G}$-bounded operators (described in Section 2.2). Part B  implies that for a large class of quantum channels (operations) this condition is also necessary for
existence of such extension.\pagebreak

\begin{theorem}\label{main}\emph{Let $\,\Phi:\B(\H)\rightarrow\B(\H)$ be a CP normal linear map s.t. $\,\Phi(I_{\H})\leq I_{\H}\,$ and $G$ a positive semidefinite densely defined operator on $\H$ satisfying condition (\ref{G-cond}).}
\medskip

\noindent A) \emph{If the map $\,\Phi$ is $G$-limited (Def.\ref{ELC}) then it is continuous on $\,\B(\H)$ w.r.t. the norm $\,\|\cdot\|_E^G$ for any $E>0$. Moreover,  any Stinespring representation (\ref{St-rep}) of $\,\Phi$  defines a unique
bounded linear operator (also denoted by $\Phi$) on the Banach space $\B_{\!G}(\H)$ of $\sqrt{G}$-bounded operators such that
\begin{equation}\label{phi-norm}
\|\Phi(A)\|_E^G\leq \sqrt{\|\Phi(I_{\H})\|}\|A\|_{Y_{\Phi}(E)}^G\leq\sqrt{\max\left\{1,Y_{\Phi}(E)/E\right\}\|\Phi(I_{\H})\|}\|A\|_{E}^G,
\end{equation}
for any $A\in\B_{\!G}(\H)$ and $E>0$, where $\,Y_{\Phi}(E)$ is the function defined in (\ref{E-phi}). The operator $\Phi$ is bounded w.r.t. the seminorm $b_{\sqrt{G}}(\cdot)$:
\begin{equation}\label{b-norm}
b_{\sqrt{G}}(\Phi(A))\leq \sqrt{k_{\Phi}\|\Phi(I_{\H})\|}\, b_{\sqrt{G}}(A)\quad \textrm{for any}\quad A\in\B_{\!G}(\H),
\end{equation}
where $k_{\Phi}=\lim_{E\rightarrow+\infty} Y_{\Phi}(E)/E$, in particular, $\Phi$ maps the subspace $\B^0_{\!G}(\H)$ of\break $\sqrt{G}$-infinitesimal operators into itself.}\footnote{The concavity of the function $E\mapsto Y_{\Phi}(E)$ implies that the nonnegative function $E\mapsto Y_{\Phi}(E)/E$ is non-increasing and hence has a finite limit as $E\rightarrow+\infty$.} \medskip

\noindent B) \emph{If $\,G$ is a discrete operator (Def.\ref{D-H}) and  one of following conditions holds}

\begin{enumerate}[\indent a)]
  \item \emph{the map $\Phi$ has finite Choi rank;}
  \item \emph{$\Phi(I_{\H})=\Phi(P)$ for some finite rank projector $P$},
\end{enumerate}
\emph{then  continuity of $\,\Phi$ on $\,\B(\H)$ w.r.t. the norm $\,\|\cdot\|_E^G$ for some $E>0$ implies
that the map $\,\Phi$ is $G$-limited.}
\end{theorem}\smallskip

\begin{remark}\label{main-r+} Since the continuity of the map $\,\Phi$ on $\,\B(\H)$ w.r.t. the norm $\,\|\cdot\|_E^G$
is necessary for existence of the bounded linear extension of $\Phi$ to $\B_{\!G}(\H)$ mentioned in part A of Theorem \ref{main}, this theorem shows
that the existence of such extension \emph{is equivalent} to the $G$-limited property  for CP linear maps
satisfying one of the conditions in part B (provided that $G$ is a discrete operator).
\smallskip
\end{remark}

\begin{remark}\label{main-r} Condition $\rm a)$ in part B of Theorem \ref{main} means that
the map $\Phi$ has the Kraus representation $\,\Phi(A)=\sum_k V^*_kA V_k\,$ with a finite number of summands. \smallskip

Condition $\rm b)$ means that the image of $\T(\H)$ under the predual map $\Phi_*$ is contained in
$\T(\H_P)$, where $\H_P$ is the finite-dimensional range of $P$.\smallskip

Conditions $\rm a)$ and $\rm b)$ are complementary to each other: if $\Phi$ satisfies condition $\rm a)$
then the complementary map $\widehat{\Phi}$ satisfies condition $\rm b)$ and vice versa.\footnote{If $\Phi$ has the Stinespring representation (\ref{St-rep}) then
the complementary map $\widehat{\Phi}$ is defined as $\,\widehat{\Phi}(A)=V_{\Phi}^*(I_{\H}\otimes A)V_{\Phi}$, $A\in\B(\K)$ \cite[Ch.6]{H-SCI}. Assuming that $\K$ is a subspace of $\H$ we may consider $\widehat{\Phi}$ as a map from $\B(\H)$ into itself.}\smallskip
\end{remark}

\emph{Proof of Theorem \ref{main}.} A) To prove continuity of $\,\Phi$ on $\,\B(\H)$ w.r.t. the norm $\,\|\cdot\|_E^G$
it suffices to assume the $2$-positivity of $\,\Phi$. Indeed, Kadison's inequality (\ref{Kad}) implies that
$$
\Tr [\Phi(A)]^*\Phi(A)\rho\leq\|\Phi(I_{\H})\|\Tr\Phi(A^*A)\rho=\|\Phi(I_{\H})\|\Tr A^*A\Phi_*(\rho)\leq \|\Phi(I_{\H})\| \left[\|A\|^{G}_{Y_{\Phi}(E)}\right]^2
$$
for any $A\in\B(\H)$ and any $\rho\in\S(\H)$ such that $\Tr {G}\rho\leq E$ (since the condition $\,\Phi(I_{\H})\leq I_{\H}\,$ guarantees  that $\Tr\Phi_*(\rho)\leq 1)$. Hence, by using inequality (\ref{E-n-eq}) we obtain
$$
\|\Phi(A)\|_E^G\leq\sqrt{\|\Phi(I_{\H})\|}\|A\|_{Y_{\Phi}(E)}^G\leq\max\left\{1,\sqrt{Y_{\Phi}(E)/E}\right\}\sqrt{\|\Phi(I_{\H})\|}\|A\|_E^G.
$$

Let $A$ be an arbitrary operator in $\,\B_{\!G}(\H)$. If the map $\,\Phi$ has representation (\ref{St-rep}) then the predual map has the form $\Phi_*(\rho)=\mathrm{Tr}_{\K}V_{\Phi}\rho V_{\Phi}^{*}$. So, the finiteness
of $Y_{\Phi}(E)$ shows that $\,V_{\Phi}|\varphi\rangle\in\D(\sqrt{G}\otimes I_{\K})\,$ for any $\,\varphi\in\D(\sqrt{G})$.  By Lemma \ref{vbl} the operator $A\otimes I_{\K}$ has a unique extension
to the set $\D(\sqrt{G}\otimes I_{\K})$ satisfying (\ref{s-prop}). So, the operator $V^*_{\Phi}[A\otimes I_{\K}]V_{\Phi}$ is well defined on  $\D(\sqrt{G})$. It does not depend on representation (\ref{St-rep}). Indeed, for any other Stinespring operator $\,V'_{\Phi}:\H\rightarrow\H\otimes\K'$
there is a partial isometry $W:\K\rightarrow\K'$ such that $V'_{\Phi}=(I_{\H}\otimes W) V_{\Phi}$ and $V_{\Phi}=(I_{\H}\otimes W^*) V'_{\Phi}$ \cite[Ch.6]{H-SCI}. So, by  Remark \ref{w-c} we have  $V^*_{\Phi}[A\otimes I_{\K}]V_{\Phi}|\varphi\rangle=[V'_{\Phi}]^*[A\otimes I_{\K}]V'_{\Phi}|\varphi\rangle$ for any $\varphi\in\D(\sqrt{G})$.

Let $\{\varphi_k\}$ be a  set of vectors in $\H$ such that $\sum_k\|\varphi_k\|^2\leq 1$ and $\;\sum_k\|\sqrt{G}\varphi_k\|^2\leq E$.
Since $V^*_{\Phi}V_{\Phi}=\Phi(I_{\H})$, we have $\|V_{\Phi}\|^2=\|\Phi(I_{\H})\|\leq 1$ and hence $\sum_k\|V_{\Phi}\varphi_k\|^2\leq 1$.
Let $\rho=\sum_k |\varphi_k\rangle\langle\varphi_k|$ be an operator in the unit ball of $\T_+(\H)$. Then $\,\Tr G \rho=\sum_k\|\sqrt{G}\varphi_k\|^2\leq E$ and hence $\,\sum_k\|\sqrt{G}\otimes I_{\K}\shs V_{\Phi}\varphi_k\|^2=\Tr G \Phi_*(\rho)\leq Y_{\Phi}(E)$. So, expression (\ref{en-def-r-1})  implies that
\begin{equation*}%\label{exp-rel}
\sum_k\|V^*_{\Phi}[A\otimes I_{\K}]V_{\Phi}\varphi_k\|^2\leq \|\Phi(I_{\H})\|\sum_k\|[A\otimes I_{\K}]V_{\Phi}\varphi_k\|^2\leq\|\Phi(I_{\H})\|\left[\|A\otimes I_{\K}\|^{G\otimes I_{\K}}_{Y_{\Phi}(E)}\right]^2.
\end{equation*}
By Lemma \ref{vbl}  the r.h.s. of this inequality coincides with $\|\Phi(I_{\H})\|\left[\|A\|^{G}_{Y_{\Phi}(E)}\right]^2$. Thus, inequality (\ref{phi-norm}) follows from expression (\ref{en-def-r-1}) for $\|V^*_{\Phi}[A\otimes I_{\K}]V_{\Phi}\|_E^G$ and inequality (\ref{E-n-eq}).\smallskip

Inequality (\ref{b-norm}) is proved by using  formula (\ref{G-bound}) and the first inequality in (\ref{phi-norm}). \smallskip

B) Assume that
\begin{equation}\label{b-rel}
\|\Phi(A)\|_E^G\leq k_E\|A\|_E^G,\quad \forall A\in\B(\H),
\end{equation}
for some $E>0$ and $k_E\in(0,+\infty)$. It follows from (\ref{E-n-eq}) that the boundedness relation (\ref{b-rel}) with finite $k_E$ holds for all $E>0$.

Assume that the operator $G$ has representation (\ref{H-rep}). Denote by $\H_n$ and $P_n$ the subspace spanned by the eigenvectors $\,\tau_1,\tau_2,...,\tau_{n-1}\,$ and the projector
on this subspace correspondingly. Let (\ref{St-rep}) be the minimal Stinespring representation for $\Phi$.
It means that
\begin{equation}\label{S-min}
\mathrm{lin}\left\{(A\otimes I_{\K}) V_{\Phi}|\varphi\rangle\,|\,A\in\B(\H), \varphi\in\H \right\}\;\textup{ is dense in }\;\H\otimes\K,
\end{equation}
where $\mathrm{lin}$ denotes the linear span of a subset of $\H\otimes\K$.

Assume that $\Phi$ has finite Choi rank, i.e. $\dim\K<+\infty$. Let $A$ be any operator in the unit ball $\B_1(\H)$ of $\B(\H)$.
Since $A\sqrt{G}P_n\in\B(\H)$ for all $n$ and $\|A\sqrt{G}P_n\|_E^G\leq\|A\|\|\sqrt{G}P_n\|_E^G\leq \sqrt{E}$, it follows from (\ref{b-rel}) that
\begin{equation}\label{one}
\|V_{\Phi}^*(A\otimes I_{\K}) (\sqrt{G}P_n\otimes I_{\K}) V_{\Phi}\varphi \|=\|V_{\Phi}^*(A\sqrt{G}P_n\otimes I_{\K})V_{\Phi}\varphi \|\leq k_{E}\sqrt{E}
\end{equation}
for  any unit vector $\varphi\in\D(\sqrt{G})$ such that $\|\sqrt{G}\varphi\|^2\leq E$.

It follows from (\ref{S-min}) that the function $\,F_{V_{\Phi}^*}(\psi)=\sup_{A\in\B_1(\H)}\|V_{\Phi}^*(A\otimes I_{\K})\psi\|\,$
does not take zero values on the unit sphere in   $\H\otimes\K$. Indeed, otherwise there is a unit vector $\psi$
in $\H\otimes\K$ such that $\langle\psi|A\otimes I_{\K} V_{\Phi}|\varphi\rangle=0$ for all $A\in\B_1(\H)$ and $\varphi\in\H$.
Thus, Lemma \ref{bl} below and (\ref{one}) imply that
\begin{equation*}%\label{two}
\Tr GP_n\Phi_*(|\varphi\rangle\langle\varphi|)=\|(\sqrt{G}P_n\otimes I_{\K})V_{\Phi}\varphi \|^2\leq \delta^{-2}k^2_{E}E
\end{equation*}
for  any unit vector $\varphi\in\D(\sqrt{G})$ such that $\|\sqrt{G}\varphi\|^2\leq E$, all $n$ and some $\delta>0$. Hence, $\Tr G\Phi_*(|\varphi\rangle\langle\varphi|)\leq \delta^{-2}k^2_{E}E\,$ for any such $\varphi$. By Proposition \ref{el-p} the map $\Phi$ is $G$-limited.\smallskip

Assume that $\Phi(P)=\Phi(I_{\H})$, where $P$ is a finite rank projector in $\B(\H)$. We will show first that
(\ref{b-rel}) implies that the vector $V_{\Phi}|\varphi\rangle$ belongs to the set $\D(\sqrt{G}\otimes I_{\K})$ for any  $\varphi\in\D(\sqrt{G})$.

Suppose, there is a vector $\varphi\in\D(\sqrt{G})$ such that $V_{\Phi}|\varphi\rangle$ does not lie in $\D(\sqrt{G}\otimes I_{\K})$.
It means that
$$
\sum_{k=0}^{+\infty} E_k \sum_{j}|\langle\tau_k\otimes \iota_j|V_{\Phi}|\varphi\rangle|^2=+\infty,
$$
where $\{ \iota_j\}$ is an orthonormal basic in $\K$.

It is easy to construct a sequence $\{c_k\}\subset(0,1)$ vanishing as $k\rightarrow+\infty$ such that
$$
\sum_{k=0}^{+\infty} c_kE_k \sum_{j}|\langle\tau_k\otimes \iota_j|V_{\Phi}|\varphi\rangle|^2=+\infty.
$$
It follows that the vector $V_{\Phi}|\varphi\rangle$ does not belong to the domain of the operator $B\otimes I_{\K}$, where
$$
B=\sum_{k=0}^{+\infty} \sqrt{c_kE_k} |\tau_k\rangle\langle\tau_k|
$$
is a positive densely defined operator on $\H$. Note that $\|B\|_E^G=o(\sqrt{E})$ as $E\rightarrow+\infty$. Indeed, since
$$
[\|B\|_E^G]^2=\sup\left\{\sum_{k=0}^{\infty} c_kE_k x_k^2\,\left|\,\sum_{k=0}^{\infty} E_k x_k^2= E, \sum_{k=0}^{\infty}x_k^2=1\right.\right\},
$$
it is easy to show that the assumption $\,\|B\|_E^G\geq C \sqrt{E}\,$ for some $\,C>0\,$ and all sufficiently large $E$ leads to a contradiction.

Thus,  $B\in\B_G^0(\H)$ and hence the sequence $\{BP_n\}\subset\B(\H)$ tends to the operator $B$ w.r.t. the norm $\|\cdot\|_E^G$ \cite[Remark 7]{ECN}.
Thus, $\{BP_n\}$ is a Cauchy sequence w.r.t. the norm $\|\cdot\|_E^G$.

Let $P_{n,m}=P_m-P_n$ for any $m>n$ and $E=\|\sqrt{G}\varphi\|^2$. It follows from (\ref{b-rel}) that
$$
\|V_{\Phi}^*(AB P_{n,m}\otimes I_{\K})V_{\Phi}\|_E^G=\|\Phi(AB P_{n,m})\|_E^G\leq k_E\|ABP_{n,m}\|_E^G\leq k_E\|BP_{n,m}\|_E^G,
$$
for any $A$ in $\B_1(\H)$. Thus,
\begin{equation}\label{imp-in}
\sup_{A\in\B_1(\H)}\|V_{\Phi}^*(A\otimes I_{\K}) (B\otimes I_{\K})\psi_{n,m}\|\leq k_E\|BP_{n,m}\|_E^G,
\end{equation}
where $|\psi_{n,m}\rangle=(P_{n,m}\otimes I_{\K})V_{\Phi}|\varphi\rangle$.

Since $\Phi(P)=\Phi(I_{\H})$, where $P$ is a projector such that $\rank P=d<+\infty$, the vector $V_{\Phi}|\varphi\rangle$ lies in $\H_d\otimes\K$, where $\H_d=P(\H)$ is a $d$-dimensional subspace of $\H$. It follows that $V_{\Phi}|\varphi\rangle=\sum_{i=1}^d |\alpha_i\rangle\otimes|\beta_i\rangle$, where $\{\alpha_i\}$ and $\{\beta_i\}$ are orthogonal sets of vectors in $\H_d$ and $\K$ correspondingly. We may assume that $\|\beta_i\|=1$ for all $i$.

For given $n$ and $m$ denote by $\H_{n,m}$ the image of the subspace $\H_d$ under the projector $P_{n,m}$. Let $U_{n,m}$ be any unitary operator
in $\B(\H)$ such that $U_{n,m}(\H_{n,m})\subseteq\H_d$. Then all the vectors
$\|(U_{n,m}B\otimes I_{\K})\psi_{n,m}\|^{-1}|(U_{n,m}B\otimes I_{\K})\psi_{n,m}\rangle$ belong to the set
$$
\A_{\beta_1,..,\beta_d}\doteq\left\{\left.\sum_{i=1}^d |\eta_i\rangle\otimes|\beta_i\rangle\,\right|\, \{\eta_i\}\subset\H_d, \;\sum_{i=1}^d \|\eta_i\|^2=1  \right\}.
$$
The set $\A_{\beta_1,..,\beta_d}$ is compact, since it is the image of the unit sphere in the direct sum of $d$ copies of $\H_d$ under the continuous map
$$
(\eta_1,...\eta_d)\mapsto \sum_{i=1}^d |\eta_i\rangle\otimes|\beta_i\rangle.
$$

The function $F_{V_{\Phi}^*}(\psi)=\sup_{A\in\B_1(\H)}\|V_{\Phi}^*(A\otimes I_{\K})\psi\|$ is lower semicontinuous on $\H\otimes\K$ (as the least upper bound of the family of continuous functions $\varphi\mapsto\|V_{\Phi}^*(A\otimes I_{\K})\psi\|$, $A\in\B_1(\H)$). So, this function attains its infimum
on the compact set $\A_{\beta_1,..,\beta_d}$ at some
vector $\psi_0\in \A_{\beta_1,..,\beta_d}$. It was mentioned before that (\ref{S-min}) implies that the function $F_{V_{\Phi}^*}$
is not equal to zero on the unit sphere in $\H\otimes\K$. Thus,
$$
\inf\{F_{V_{\Phi}^*}(\psi)\,|\,\psi\in \A_{\beta_1,..,\beta_d}\}=F_{V_{\Phi}^*}(\psi_0)>0.
$$
Denoting this infimum by $\delta$, we obtain
\begin{equation}\label{imp-in+}
\begin{array}{rl}
F_{V_{\Phi}^*}((B\otimes I_{\K})\psi_{n,m})\!\! & =F_{V_{\Phi}^*}((U_{n,m}B\otimes I_{\K})\psi_{n,m})\\\\
& \geq \delta\|(U_{n,m}B\otimes I_{\K})\psi_{n,m}\|=\delta\|(B\otimes I_{\K})\psi_{n,m}\|\quad \forall n,m,
\end{array}
\end{equation}
where the first equality follows from the definition of the function $F_{V_{\Phi}^*}$.

Since $\{BP_n\}$ is a Cauchy sequence in $\B(\H)$ w.r.t. the norm $\|\cdot\|_E^G$, the r.h.s. of (\ref{imp-in}) tends to zero
as $n,m \rightarrow+\infty$. Thus, (\ref{imp-in}) and (\ref{imp-in+}) imply that $\{|(B\otimes I_{\K})\psi_{n}\rangle\}$, where $|\psi_{n}\rangle=(P_{n}\otimes I_{\K})V_{\Phi}|\varphi\rangle$, is a Cauchy sequence of vectors in $\H\otimes\K$. So, it has a limit in $\H\otimes\K$. Since the sequence $\{|\psi_{n}\rangle\}$ tends to the vector $V_{\Phi}|\varphi\rangle$
and the operator $B\otimes I_{\K}$ is closed, the vector $V_{\Phi}|\varphi\rangle$ belongs to the domain of $B\otimes I_{\K}$ contradicting the
above assumption.  Thus,
\begin{equation}\label{b-inc}
  V_{\Phi}|\varphi\rangle\in\D(\sqrt{G}\otimes I_{\K})\;\textup{ for any }\;\varphi\in\D(\sqrt{G}).
\end{equation}

The density of $\D(\sqrt{G})$ in $\H$ shows that the minimal subspace $\H_d$ containing the supports of all the states $\Tr_{\K} V_{\Phi}\rho V_{\Phi}^*$, $\rho\in\S(\H)$, is spanned by the eigenvectors of all the states
$\Tr_{\K} V_{\Phi}|\varphi\rangle\langle\varphi|V_{\Phi}^*$, $\varphi\in\D(\sqrt{G})$, corresponding to nonzero
eigenvalues. By using this it is easy to show that (\ref{b-inc}) implies that $\H_d\subset\D(\sqrt{G})$. It follows that the restriction of $\sqrt{G}$ to the subspace $\H_d$ is a bounded operator, and hence the restriction of $\sqrt{G}\otimes I_{\K}$ to the subspace
$\H_d\otimes\K$ is a bounded operator. Thus, $$
\varphi\mapsto\Tr G\Phi_*(|\varphi\rangle\langle\varphi|)=\|\sqrt{G}\otimes I_{\K}V_{\Phi}\varphi\|^2
$$
is a bounded function on the unit ball of $\H$. It follows that $\rho\mapsto\Tr G\Phi_*(\rho)$ is a bounded function on $\S(\H)$. $\square$ \smallskip

\begin{lemma}\label{bl}
\emph{Let $\H$ and $\K$ be separable Hilbert space and $\,\dim\K<+\infty$. Let $C$ be any operator  in $\B(\H\otimes\K)$. If
$$
F_C(\varphi)\doteq\sup_{A\in\B_1(\H)} \|C (A\otimes I_{\K})\varphi\|>0\;\textit{ for all }\;\varphi\neq 0\; \textit{ then }\; \inf_{\varphi\in[\H\otimes\K]_1} F_C(\varphi)>0
$$
where $\B_1(\H)$ is the unit ball in  $\B(\H)$ and $[\H\otimes\K]_1$ is the unit sphere in $\H\otimes\K$, and this infimum is attainable.}
\end{lemma}\smallskip

\begin{remark}\label{bl-r} The assertion of Lemma \ref{bl} is not valid if $\,\dim\K=+\infty$ \cite{Shulman}.
\end{remark}\smallskip

\emph{Proof.} Note first that the function $F_C(\varphi)$ is lower semicontinuous on $\H\otimes\K$ (as the least upper bound of the family of continuous functions $\varphi\mapsto\|C(A\otimes I_{\K})\varphi\|$, $A\in\B(\H)$.

Let $\{\tau_k\}$ be an orthonormal basis in $\H$, $n=\dim\K$ and $\A$ be the set of unit vectors $\H\otimes\K$ having the representation
$$
|\varphi\rangle=\sum_{k=1}^n |\tau_k\rangle\otimes|\psi_k\rangle, \quad \{\psi_k\}\subset\K,\;\; \sum_{k=1}^n\|\psi_k\|^2=1.
$$
Since $\A$ is the image of the unit sphere in the direct sum of $n$ copies of $\K$ under the continuous map
$$
(\psi_1,...\psi_n)\mapsto \sum_{k=1}^n |\tau_k\rangle\otimes|\psi_k\rangle,
$$
the set $\A$ is compact. So, the lower semicontinuous function $F_C(\varphi)$ attains its infimum on $\A$ at some
vector $\varphi_0\in \A$ and hence this infimum is positive by the assumption.

To complete the proof of the lemma it suffices to note that
$$
\inf_{\varphi\in[\H\otimes\K]_1} F_C(\varphi)=F_C(\varphi_0).
$$
This follows from the definition of $F_C$, since any vector in $[\H\otimes\K]_1$ can be represented as $(U\otimes I_{\K})|\varphi\rangle$, where $U$ is a unitary operator in $\B(\H)$ and $\varphi$ is a vector in $\A$. $\square$
\bigskip

\section{Generalizations to quantum channels and operation between different systems}

For simplicity in the previous sections we restrict attention to CP linear trasformations of the algebra $\B(\H)$. All the results presented therein are directly generalized to CP normal linear maps from $\B(\H)$ into $\B(\H')$, where $\H$ and $\H'$  are separable Hilbert spaces. Since all such spaces are isomorphic, the only essential feature of this case is the possibility to chose \emph{different} operators $G$ and $G'$ on these spaces.
So, we have to generalize the notion of $G$-limited quantum operation as follows.

\begin{definition}\label{ELC+} A quantum operation $\Phi:\B(\H)\rightarrow\B(\H')$ is called $GG'$-limited if
\begin{equation}\label{E-phi+-}
\Tr G\Phi_*(\rho)<+\infty\;\textrm{ for any }\;\rho\in\S(\H')\;\textrm{ such that }\;\Tr G'\rho<+\infty,
\end{equation}
where $\Phi_*:\T(\H')\rightarrow\T(\H)$ is the predual operation.
\end{definition}\smallskip

By Lemma \ref{sl}  condition (\ref{E-phi+-}) is equivalent to the following
\begin{equation}\label{E-phi+-+}
Y_{\Phi}(E)\doteq \sup\left\{\shs\Tr {G}\Phi_*(\rho)\,|\,\rho\in\mathfrak{S}(\H'),\Tr {G'}\rho\leq E\,\right\}<+\infty
\end{equation}
for some $E>0$. Since the function $\,E\mapsto Y_{\Phi}(E)\,$ is concave,
the finiteness of $Y_{\Phi}(E)$ for some $E>0$ implies its finiteness for all $E>0$.\smallskip

The following proposition generalizes Proposition \ref{el-p} in Section 2.3. It is proved similarly by using Lemma \ref{nbl}.\smallskip

\begin{property}\label{el-p+}
\emph{If  $G'$ is a positive unbounded discrete operator (\ref{H-rep}) then
$$
Y_{\Phi}(E)\doteq \sup\left\{\shs\Tr {G}\Phi_*(|\varphi\rangle\langle\varphi|)\,\left|\, \varphi\in\H_*, \|\sqrt{G'}\varphi\|^2\leq E, \|\varphi\|=1\!\right.\right\},
$$
for any quantum operation $\,\Phi:\B(\H)\rightarrow\B(\H')$, where $\H'_*$ is the linear span of all the eigenvectors $\,\tau_1,\tau_2,...\,$ of $G'$, i.e. the supremum in (\ref{E-phi+-+}) can be taken only over pure states corresponding to the vectors in $\H'_*$.}
\end{property}\smallskip

Note that any CP normal linear map $\,\Phi:\B(\H)\rightarrow\B(\H')$ has the Stinespring representation (\ref{St-rep}), where $V_{\Phi}$ is a linear operator from $\H'$ into $\H\otimes\K$. \smallskip

\begin{theorem}\label{main+}\emph{Let $\,\Phi:\B(\H)\rightarrow\B(\H')$ be a normal CP linear map s.t. $\,\Phi(I_{\H})\leq I_{\H'}\,$. Let $G$ and $G'$ be positive semidefinite operators densely defined, respectively, on $\H$ and $\H'$ satisfying condition (\ref{G-cond}).}
\medskip

\noindent A) \emph{If the map $\,\Phi$ is $GG'$-limited (Def.\ref{ELC+}) then it is continuous w.r.t. the norms $\,\|\cdot\|_E^G$ and $\,\|\cdot\|_E^{G'}$ on $\,\B(\H)$ and $\,\B(\H')$ for any $E>0$. Moreover,  any Stinespring representation (\ref{St-rep}) of $\,\Phi$  defines a unique
bounded linear operator (also denoted by $\,\Phi$) from $\B_{\!G}(\H)$ into $\B_{\!G'}(\H')$ such that
\begin{equation}\label{phi-norm-g}
\|\Phi(A)\|_E^{G'}\leq \sqrt{\|\Phi(I_{\H})\|}\|A\|_{Y_{\Phi}(E)}^G\leq \sqrt{\max\left\{1,Y_{\Phi}(E)/E\right\}\|\Phi(I_{\H})\|}\|A\|_{E}^G
\end{equation}
for any $A\in\B_{\!G}(\H)$ and $E>0$, where $\,Y_{\Phi}(E)$ is the function defined in (\ref{E-phi+-+}).
The operator $\Phi$ is bounded w.r.t. the seminorms $b_{\sqrt{G}}(\cdot)$ and $b_{\sqrt{G'}}(\cdot)$ on $\B_{\!G}(\H)$ and $\B_{\!G'}(\H')$:
$$
b_{\sqrt{G'}}(\Phi(A))\leq \sqrt{k_{\Phi}\|\Phi(I_{\H})\|}\, b_{\sqrt{G}}(A)\quad \textrm{for any}\quad A\in\B_{\!G}(\H),
$$
where $\,k_{\Phi}=\lim_{E\rightarrow+\infty} Y_{\Phi}(E)/E$, in particular, $\Phi$ maps  $\B^0_{\!G}(\H)$ into
 $\B^0_{\!G'}(\H')$.}\medskip

\noindent B) \emph{If $\,G'$ is a discrete operator (Def.\ref{D-H}) and  one of following conditions holds}

\begin{enumerate}[\indent a)]
  \item \emph{the map $\Phi$ has finite Choi rank;}
  \item \emph{$\Phi(I_{\H})=\Phi(P)$ for a finite rank projector $P\in\B(\H)$ and $\,G$ is a discrete operator}
\end{enumerate}
\emph{then  continuity of $\,\Phi$ w.r.t. the norms $\,\|\cdot\|_E^G$ and $\,\|\cdot\|_E^{G'}$ on $\,\B(\H)$ and $\,\B(\H')$  for some $E>0$ implies
that the map $\,\Phi$ is $GG'$-limited.}
\end{theorem}\smallskip

Theorem \ref{main+} is proved by obvious modifications of the arguments from the proof of Theorem \ref{main}.
\smallskip

\begin{remark}\label{main+r}
Since continuity of a map $\,\Phi:\B(\H)\rightarrow\B(\H')$ w.r.t. the norms $\,\|\cdot\|_E^G$ and $\,\|\cdot\|_E^{G'}$ 
is necessary for existence of the bounded linear extension of $\Phi$ to $\B_{\!G}(\H)$ mentioned in part A of Theorem \ref{main+}, part B of this theorem shows
that the existence of such extension implies  the $GG'$-limited property of a map $\Phi$ (provided that one of the conditions a) and b) holds).
\end{remark}

\section{Bosonic Gaussian Channels}

In this section we apply Theorem \ref{main+} to the class of Bosonic Gaussian channels
playing central role in the modern quantum information theory \cite{H-SCI,W&C}.\smallskip

Let $\mathcal{H}$ and $\mathcal{H}'$ be the spaces of irreducible representation of
the Canonical Commutation Relations (CCR)
\begin{equation*}
W(z_1)W(z_2)=\exp
\left(-\textstyle{\frac{\mathrm{i}}{2}}\,z_1^{\top }\Delta
z_2\right) W(z_1+z_2),\;\quad\; z_1,z_2\in Z,
\end{equation*}
\begin{equation*}
W'(z_1)W'(z_2)=\exp
\left(-\textstyle{\frac{\mathrm{i}}{2}}\,z_1^{\top }\Delta'
z_2\right) W'(z_1+z_2),\quad z_1,z_2\in Z'\!\!,
\end{equation*}
where  $(Z,\Delta)$ and $(Z',\Delta')$ are symplectic spaces,  $\{W(z)\}_{z\in Z}$ and
$\{W'(z)\}_{z\in Z'}$ are the families of Weyl operators in $\B(\H)$ and $\B(\H')$ correspondingly \cite[Ch.12]{H-SCI}. Denote by $s$ and $s'$ the
numbers of modes of the systems, i.e. $2s=\dim Z$ and $2s'=\dim Z'$.

A Bosonic Gaussian channel  $\Phi_{K,\alpha,\ell}
:\mathfrak{B}(\H)\rightarrow
\mathfrak{B}(\H')$ is defined via the action on the Weyl operators:
\begin{equation}
\Phi_{K,\alpha,\ell}(W(z))=W'(Kz)\exp \left[\, \mathrm{i}\hspace{1pt}\ell\hspace{1pt}z-\textstyle\frac{1}{2}\hspace{1pt}z^{\top }\alpha
\hspace{1pt}z\,\right],\quad z\in Z, \label{blc}
\end{equation}
where $K:Z\rightarrow Z'$ is a linear operator,  $\ell\,$ is a $\,2s$-dimensional real row and $\,\alpha\,$ is a
real symmetric $\,(2s)\times(2s)$ matrix satisfying the
inequality $\alpha \geq \pm \frac{\mathrm{i}}{2}\left[ \Delta-K^{\top }\Delta' K\right]$ \cite{E&W,H-SCI}.

Any such channel can be transformed by appropriate displacement
unitaries to the Bosonic Gaussian channel with $\,\ell=0\,$ and the
same matrix $\,\alpha\,$ \cite{H-SCI}. So, we will restrict attention to Gaussian  channels $\Phi_{K,\alpha,0}$
typically called \emph{centred} and denoted by $\Phi_{K,\alpha}$ in what follows.

Let $R=[q_1,p_1,...,q_s,p_s]^{\top}$ be the $2s$-vector of the canonical observables -- linear unbounded operators on the space $\H$ with common dense domain satisfying the commutation relations
$$
[q_i,p_j]=\mathrm{i}\delta_{ij}I_{\H},\quad [q_i,q_j]=[p_i,p_j]=0\quad \forall i,j.
$$
Physically, the operators $\,q_1,...,q_s\,$ and $\,p_1,...,p_s\,$ can be treated, respectively, as generalized position and  momentum observables of the $s$-mode quantum oscillator.

Let $R'=[q'_1,p'_1,...,q'_{s'},p'_{s'}]^{\top}$ be the $2s'$-vector of the canonical observables of the Bosonic system described by the space $\H'$.

Any Gaussian channel $\,\Phi_{K,\alpha}:\B(\H)\rightarrow\B(\H')\,$ can be correctly extended  to the algebra of all
polynomials in the
canonical observables $\,q_1,...,q_s\,$ and $\,p_1,...,p_s\,$. Moreover, it transforms any such polynomial into a polynomial in the canonical observables $\,q'_1,...,q'_{s'}\,$ and $\,p'_1,...,p'_{s'}\,$ of the same order.  This property can be proved by differentiating the relation (\ref{blc}) at the point $\,z=0\,$ with the help of Baker-Campbell-Hausdorff formula \cite{H-SCI}. In particular, by using this
way one can show  that
\begin{equation}\label{imp-e}
\Phi_{K,\alpha}(R^{\top}R)=\Phi_{K,\alpha}\!\left(\sum_{i=1}^s q_i^2+p_i^2\right)=[R']^{\top}K^{\top}KR'+I_{\H'}\mathrm{Sp}\alpha,
\end{equation}
where $\mathrm{Sp}$ denotes the spur (trace) of a $\,(2s)\times(2s)$ matrix.\smallskip

In many cases the Hamiltonian (energy observable) of a Bosonic system has the form $R^{\top}\varepsilon R$, where  $\varepsilon$ is a real positive nondegenerate matrix \cite{H-SCI,W&C}. So, we will assume that
\begin{equation}\label{G-q-form}
G=R^{\top}\varepsilon R-c_{\varepsilon} I_{\H}\quad \textrm{and} \quad  G'=[R']^{\top}\varepsilon' R'-c_{\varepsilon'} I_{\H'},
\end{equation}
where $\varepsilon$ and $\varepsilon'$ are real positive nondegenerate $\,(2s)\times(2s)$ and $\,(2s')\times(2s')$ matrices,
$c_{\varepsilon}$ and $c_{\varepsilon'}$ are the infima of the
spectrum of the positive operators $R^{\top}\varepsilon R$ and $[R']^{\top}\varepsilon' R'$ correspondingly.  In the following lemma we use the notion of a $GG'$-limited channel (Def.\ref{ELC+}) and
the function $Y_{\Phi}(E)$ defined in (\ref{E-phi+-+}).\smallskip

\begin{lemma}\label{Guass-l} \emph{Let $G$ and $G'$ be the operators defined in (\ref{G-q-form}). Then any Gaussian channel $\,\Phi_{K,\alpha}:\B(\H)\rightarrow\B(\H')\,$ is $GG'$-limited and
\begin{equation}\label{Y-b}
Y_{\Phi_{K,\alpha}}(E)\leq aE+ac_{\varepsilon'}-c_{\varepsilon}+b,
\end{equation}
where $\,a=\|\varepsilon\|\|K\|^2/m(\varepsilon')$, $m(\varepsilon')$ is the minimal eigenvalue of $\,\varepsilon'$, and $\,b=\|\varepsilon\|\mathrm{Sp}\alpha$.\footnote{Here $\|A\|$ denotes  the operator norm of a matrix $A$.}}\smallskip
\end{lemma}\smallskip

\emph{Proof.} By using relation (\ref{imp-e}) we obtain
$$
\begin{array}{rl}
\Phi_{K,\alpha}(R^{\top}\varepsilon R)\!\! & \leq \,\|\varepsilon\|\Phi_{K,\alpha}(R^{\top}R)\leq
\|\varepsilon\|\|K\|^2  [R']^{\top}R'+\|\varepsilon\|I_{\H'}\mathrm{Sp}\alpha\\\\
& \leq\,
\|\varepsilon\|\|K\|^2 m(\varepsilon')^{-1} [R']^{\top}\varepsilon' R'+\|\varepsilon\|I_{\H'}\mathrm{Sp}\alpha.
\end{array}
$$
Hence
$$
\Tr G[\Phi_{K,\alpha}]_*(\rho)=\Tr\shs\Phi_{K,\alpha}(G)\rho\leq a\Tr G'\rho+ac_{\varepsilon'}-c_{\varepsilon}+b
$$
for any state $\rho\in\S(\H')$.  This inequality implies (\ref{Y-b}).  $\square$ \smallskip

Speaking about extension of Gaussian channels to unbounded operators we may (w.l.o.g.) chose on the role of $G$ and $G'$ the \emph{number operators}
\begin{equation}\label{N-form}
 N=\textstyle\frac{1}{2}(R^{\top} R-sI_{\H})\quad \textrm{and}\quad N'=\textstyle\frac{1}{2}([R']^{\top} R'-s'I_{\H'}).
\end{equation}
Indeed, since $\,m(\varepsilon )R^{\top}R\leq R^{\top}\varepsilon R\leq \|\varepsilon\|R^{\top}R$, where $m(\varepsilon)$ is the minimal eigenvalue of $\,\varepsilon$, it is easy to see that $\B_{\!G}(\H)=\B_{\!N}(\H)$ for any  operator $G$ defined in (\ref{G-q-form}) and that the norms
$\,\|\cdot\|_E^G$ and $\,\|\cdot\|_E^N$ are equivalent. The same arguments show that $\B_{\!G'}(\H')=\B_{\!N'}(\H')$ for any  operator $G'$ defined in (\ref{G-q-form}) and that the norms $\,\|\cdot\|_E^{G'}$ and $\,\|\cdot\|_E^{N'}$ are equivalent.\smallskip

It is easy to see that the Banach space $\B_{\!N}(\H)$ (corresp. $\B_{\!N'}(\H')$) contains all the canonical observables $\,q_1,...,q_s\,$ and $\,p_1,...,p_s\,$ (corresp., $\,q'_1,...,q'_{s'}\,$ and $\,p'_1,...,p'_{s'}\,$) and their linear combinations.\footnote{The operators \emph{E}-norms of the observables $q$ and $p$ in the case $s=1$ are estimated in \cite[Sect.5]{ECN}.}
\smallskip

If $G=N$ and $G'=N'$ then it follows from (\ref{Y-b}) that $$Y_{\Phi_{K,\alpha}}(E)\leq F_{K,\alpha}(E)\doteq aE+(as'-s +b)/2,$$
where $\,a=\|K\|^2$, $\,b=\mathrm{Sp}\alpha$, $s=\dim Z/2$ and $s'=\dim Z'/2$. Hence Theorem \ref{main+}A implies the following\smallskip

\begin{property}\label{main-c}\emph{Let $N$ and $N'$ be the number operators (\ref{N-form}) and $\,\Phi_{K,\alpha}$ a centered Gaussian channel from $\B(\H)$ to $\B(\H')$. Then}
\begin{itemize}
  \item \emph{the channel $\,\Phi_{K,\alpha}$
is continuous w.r.t. the norms $\,\|\cdot\|_E^N$ and $\,\|\cdot\|_E^{N'}$ on $\,\B(\H)$ and $\,\B(\H')$ for any $E>0$;}
  \item \emph{any Stinespring representation  of $\,\Phi_{K,\alpha}$  defines a unique
bounded linear operator from $\B_{\!N}(\H)$ to $\B_{\!N'}(\H')$ such that \footnote{$\B_{N}(\H)$ and $\B_{N'}(\H')$ are the Banach spaces of $\sqrt{N}$-bounded operators on $\H$ and $\sqrt{N'}$-bounded operators on $\H'$ equipped with the norms $\|\cdot\|_N^E$ and $\|\cdot\|_{N'}^E$ correspondingly (see Section 2.2).}
\begin{equation}\label{G-phi-norm}
\|\Phi_{K,\alpha}(A)\|_E^{N'}\leq \|A\|_{F_{K,\alpha}(E)}^N \leq\max\left\{1,\sqrt{F_{K,\alpha}(E)/E}\right\}\|A\|_E^N, \quad \forall E>0,
\end{equation}
for any $A\in\B_{\!N}(\H)$, where $F_{K,\alpha}(E)$ is the above defined function.}\smallskip

\item \emph{$\,b_{\sqrt{N'}}(\Phi_{K,\alpha}(A))\leq \|K\|\,b_{\sqrt{N}}(A)$ for any $A\in\B_{\!N}(\H)$, in particular, $\Phi_{K,\alpha}$ maps the subspace $\B^0_{\!N}(\H)$ of $\sqrt{N}$-infinitesimal operators into the similar subspace $\B^0_{\!N'}(\H)$.}
\end{itemize}
\end{property} \smallskip

\begin{remark}\label{main-c-r} It is well known that any Gaussian channel $\Phi_{K,\alpha}$ is well defined on the linear span of the canonical observables $\,q_1,...,q_s\,$ and $\,p_1,...,p_s\,$ contained in  $\B_{N}(\H)$ and maps it into the linear span of the canonical observables $\,q'_1,...,q'_{s'}\,$ and $\,p'_1,...,p'_{s'}\,$ contained in  $\B_{N'}(\H')$. Proposition \ref{main-c} states that the channel $\Phi_{K,\alpha}$ is correctly extended to the space $\B_{N}(\H)$ of \emph{all} $\sqrt{N}$-bounded operators on $\H$ and that this extension maps $\B_{N}(\H)$ into $\B_{N'}(\H')$  continuously w.r.t. the norms $\|\cdot\|_N^E$ and $\|\cdot\|_{N'}^E$.
\end{remark}
\medskip

\textbf{Example: one-mode attenuation/amplification Gaussian channel.} Assume that $\H'=\H$, $Z=Z'$, $s=s'=1$. Consider the channel $\Phi_{K,\alpha}:\B(\H)\rightarrow\B(\H)$, where
$$
K=\left[\begin{array}{ll}
        k & 0\\
        0 & k
        \end{array}\right]\quad \textrm{and}   \quad \alpha=\left[\begin{array}{ll}
        \lambda & 0\\
        0 & \lambda
        \end{array}\right]\!,\quad\lambda=N_c+|k^2-1|/2.
$$
This is a one-mode  attenuation/amplification channel,  $\,N_c\geq0\,$ is the power of environment noise and $\,k>0\,$ is
the coefficient of attenuation/amplification \cite{H-SCI}.

In this case one can obtain explicit expression for the function $Y_{\Phi_{K,\alpha}}(E)$. Indeed, formula (\ref{imp-e}) implies that
$\Phi_{K,\alpha}(N)=k^2N+((k^2-1)/2+\lambda)I_{\H}$. Hence
$$
Y_{\Phi_{K,\alpha}}(E)=F_{k,N_c}(E)\doteq k^2E+(k^2-1)/2+\lambda=\left\{\begin{array}{ll}
        k^2E+N_c, & k\leq 1\\
        k^2E+N_c+(k^2-1), & k>1
        \end{array}\right.
$$
So, it follows from (\ref{phi-norm}) that\footnote{In this case formula (\ref{phi-norm}) coincides with formula (\ref{G-phi-norm}) obtained from (\ref{phi-norm}) by a rough upper estimate.}
\begin{equation}\label{G-phi-norm+}
\|\Phi_{K,\alpha}(A)\|_E^N\leq \|A\|_{F_{k,N_c(E)}}^N\leq \max\left\{1,\sqrt{F_{k,N_c}(E)/E}\right\}\|A\|_E^N, \qquad \forall E>0,
\end{equation}
for any $A\in\B_{\!N}(\H)$. If $\Phi_{K,\alpha}$ is a quantum-limited attenuator ($k<1$, $N_c=0$) then $\|\Phi_{K,\alpha}(A)\|_E^N\leq \|A\|_{k^2E}\leq \|A\|_E^N$ for any $A\in\B_{\!N}(\H)$, i.e.  $\Phi_{K,\alpha}$ is a contraction w.r.t. the norm $\|\cdot\|_E^N$ for any $E>0$.\smallskip

Since $\lim_{E\rightarrow+\infty}F_{k,N_c}(E)/E=k^2$ for any $k>0$ and $N_c\geq 0$,  we have
$$
b_{\sqrt{N}}(\Phi_{K,\alpha}(A))\leq k b_{\sqrt{N}}(A)\quad\textrm{for all}\quad A\in\B_{\!N}(\H).
$$
\bigskip

I am grateful to A.S.Holevo and to the participants of his seminar
"Quantum probability, statistic, information" (the Steklov
Mathematical Institute) for useful discussion. I am also grateful to G.G.Amosov,  A.V.Bulinsky and V.Zh.Sakbaev for discussion and valuable  remarks.
Special thanks to T.V.Shulman for the example showing that the condition $\dim\K<+\infty$ in Lemma \ref{bl} is essential
and to S.Weis for the idea used (implicitly) in the proof of Lemma \ref{nbl}.

\end{document}